\documentclass[conference]{IEEEtran}

\usepackage{cite}

\ifCLASSINFOpdf
  \usepackage[pdftex]{graphicx}
  \usepackage{booktabs}
\else
\fi
\begin{document}
%
\title{RealmEye: Virtual Machine Introspection for Arm CCA Realm VMs}

\author{
\IEEEauthorblockN{Ruofei Qu\IEEEauthorrefmark{1}\IEEEauthorrefmark{2},
Wei Feng\IEEEauthorrefmark{1},
Hongzhan Ma\IEEEauthorrefmark{1}\IEEEauthorrefmark{2},
Menghan Jia\IEEEauthorrefmark{1}\IEEEauthorrefmark{2},
Muyan Shen\IEEEauthorrefmark{2},
Yu Qin\IEEEauthorrefmark{1}}
\IEEEauthorblockA{\IEEEauthorrefmark{1}Institute of Software, Chinese Academy of Sciences, Beijing 100190, China}
\IEEEauthorblockA{\IEEEauthorrefmark{2}University of Chinese Academy of Sciences, Beijing 100049, China}
}
	

%


\IEEEoverridecommandlockouts
\makeatletter\def\@IEEEpubidpullup{6.5\baselineskip}\makeatother

\maketitle

\begin{abstract}

Confidential VMs (CVMs) have become the dominant substrate for sensitive cloud workloads, from financial services to privacy-preserving AI inference. The hardware isolation that protects these CVMs from a malicious cloud, however, also blinds their owners to what runs inside them: kernel rootkits planted via network or supply-chain attacks can hide processes, tamper with kernel data structures, and exfiltrate model weights, all under the cover of the same isolation that defends the VM. Cloud tenants therefore need a way to forensically inspect a running CVM from outside, yet every classical approach to doing so collapses under the CVM threat model. Traditional VM introspection (VMI) presupposes a trusted Hypervisor, which CVMs explicitly exclude from the TCB. The state-of-the-art CVM-VMI system, 00SEVen, restores introspection on AMD SEV-SNP by hosting an agent inside the VM at a privileged tier (VMPL0)---a mechanism that simply does not exist on Arm CCA, leaving Realm VMs without any introspection solution.
We present RealmEye, the first VMI system for Arm CCA Realm VMs. RealmEye places the entire introspection logic inside the Realm Management Monitor (RMM) at R-EL2, achieving hardware-enforced separation between the monitor and the monitored VM: no agent runs inside the Realm, and the Realm itself remains unmodified. From this vantage point, RealmEye reads Realm memory and registers, suspends the VM for consistent snapshots, and traps page-level accesses, all without relying on any interface the in-VM OS exposes. We extend the RMI to carry VMI triggers and encrypted results between the Host and the RMM, bind the result channel to a hardware-attested session with the remote owner, and add a CCA driver backend to LibVMI so that existing tools such as DRAKVUF interoperate with RealmEye unchanged. RealmEye is therefore isolated from the cloud platform---including a malicious Hypervisor---by Arm CCA's own hardware mechanisms, and isolated from in-Realm rootkits by the R-EL2 boundary of the RMM. A periodic, self-driven trigger mode keeps scan timing internal to the RMM, preventing the Hypervisor from colluding with in-Realm rootkits. On the Arm FVP, RealmEye detects both process hiding and syscall-table hooking by Diamorphine, a real-world ARM64 rootkit, and its in-RMM cost is linearly predictable from primitive invocation counts---laying the foundation for the first introspection ecosystem on Arm CCA.
\end{abstract}


%
\IEEEpeerreviewmaketitle

\section{Introduction}
\label{sec:intro}

Confidential VMs (CVMs) have become a primary substrate for sensitive cloud workloads, from financial services to privacy-preserving AI inference~\cite{tan2025pipellm}. Secure Encrypted Virtualization-Secure Nested Paging (SEV-SNP)~\cite{sev2020strengthening}, Intel Trust Domain Extensions (TDX)~\cite{intel2023tdx}, and Arm CCA~\cite{li2022design} all use hardware-enforced memory encryption and access control to shield a VM's memory and registers from the cloud platform. The same isolation, however, also blinds the VM owner to the VM's runtime state. The VM still runs a full OS exposed to network and supply-chain attacks, and an attacker who gains kernel control can deploy a rootkit to hide processes, tamper with kernel data, or exfiltrate sensitive data. Detecting such intrusions from outside is the classical task of Virtual Machine Introspection (VMI)~\cite{garfinkel2003virtual}, but traditional VMI presupposes a trusted Hypervisor---an assumption CVMs explicitly invalidate, since the Hypervisor is excluded from the TCB and barred by hardware from VM state. Traditional VMI thus fails on every CVM platform.

To restore VMI on CVMs, 00SEVen~\cite{schwarz202400seven} introduced an in-VM agent design on AMD SEV-SNP, hosting a hardware-protected agent at the highest VMPL tier inside the VM. This design is tightly bound to SEV-SNP-specific mechanisms (VMPL layering, RMPADJUST, SVSM); Arm CCA offers no equivalent intra-VM privilege layering, leaving no hardware-isolated tier inside a Realm in which to host such an agent. Meanwhile, CCA's Granule Protection Table (GPT) blocks the Hypervisor from Realm memory at the hardware level, ruling out Hypervisor-based VMI as well. Neither approach is applicable to Arm CCA, leaving Realm VMs without an introspection solution.

This paper presents RealmEye, the first VMI system for Arm CCA Realm VMs. RealmEye places the entire VMI logic inside the Realm Management Monitor (RMM) at R-EL2, with no agent deployed in the Realm, thereby achieving hardware-enforced separation between the monitor and the monitored VM, leaving the Realm VM untouched and fully isolated from the VMI logic. We extend the Realm Management Interface (RMI) to carry VMI triggers and encrypted results between the Host and the RMM, leaving the Hypervisor with only opaque flags and ciphertext to relay; we further bind the result channel to a hardware-attested session with the remote owner, and add a CCA driver backend to LibVMI~\cite{payne2012simplifying} so that existing tools such as DRAKVUF~\cite{lengyel2014scalability} interoperate with RealmEye unchanged. Beyond functional VMI, RealmEye also addresses two threats that CCA imposes on the introspection design itself: \emph{collusion}, where an untrusted Hypervisor signals scan timings to an in-VM rootkit, addressed by a periodic mode in which the RMM autonomously decides scan timings invisible to the Hypervisor; and \emph{symbol-input tampering}, addressed by an autonomous symbol-resolution scheme rooted at the CPU-maintained VBAR\_EL1 register, accepting no external input. On the Arm FVP, RealmEye detects process hiding and syscall-table hooking by Diamorphine, a real-world ARM64 rootkit, and our microbenchmarks and macrobenchmarks show that the cost of in-RMM VMI operations is linearly predictable from primitive invocation counts.

\noindent
This paper makes the following contributions:
\begin{itemize}
  \item We identify three CCA-specific challenges for VMI---the absence of memory-introspection interfaces in the RMM, the REC-lock concurrency conflict with VMI triggers, and Stage-2 TLB coherence---and address each with a systematic solution.
  \item We design and implement RealmEye, the first RMM-based VMI system for Arm CCA Realm VMs, placing all VMI logic in the RMM without adding any code inside the Realm.
  \item We propose two security mechanisms tailored to CCA's strengthened threat model: a periodic-trigger mode against Hypervisor--rootkit collusion, and an autonomous symbol-resolution scheme against tampered symbol input. Both substantively strengthen threat-model coverage relative to 00SEVen.
  \item We integrate RealmEye into the existing VMI ecosystem through a new CCA driver backend in LibVMI, and evaluate it on the Arm FVP with Diamorphine, characterizing its cost through micro- and macrobenchmarks.
\end{itemize}
\section{background}
\subsection{Arm CCA Architecture}

\begin{figure}[t]
  \centering
  \includegraphics[width=\columnwidth]{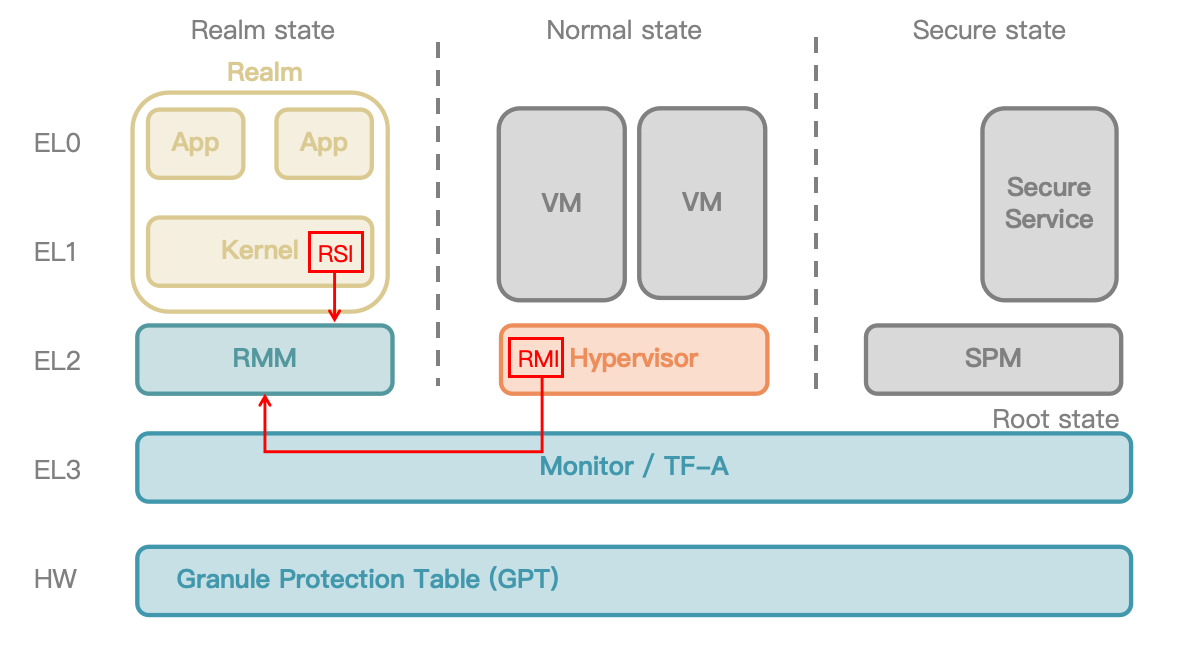}
  \caption{Arm CCA architecture overview. Adapted from~\cite{arm2023rme}.}
  \label{fig:cca-arch}
\end{figure}

Arm CCA, introduced in Armv9-A~\cite{arm2023rme}, extends the system's execution environment from the traditional two security states (Normal and Secure) into four mutually isolated worlds: Normal, Realm, Secure, and Root, as shown in Figure~\ref{fig:cca-arch}. The Normal world hosts the untrusted Hypervisor at EL2 and its conventional VMs at EL1/EL0. The Realm world hosts confidential VMs (called Realms), with the Realm Management Monitor (RMM) running at R-EL2 and the Realm VM kernel and user-space applications at R-EL1 and R-EL0. The Root world runs the EL3 Monitor firmware (TF-A), which mediates context switches between the four worlds. The Secure world hosts trusted services such as OP-TEE. Isolation between worlds is enforced by hardware, and any cross-world interaction must be relayed through the Monitor. In particular, the Hypervisor is excluded from the Realm VM's Trusted Computing Base (TCB).

CCA enforces hardware-level memory isolation between the four worlds through the Granule Protection Table (GPT), a physical-page-granularity access-control table configured by the EL3 Monitor at boot time. Each physical page (granule) has a GPT entry tagging the world it belongs to, and the hardware checks the GPT on every memory access; an unauthorized access raises a Granule Protection Fault. Realm memory is tagged as Realm-state, accessible only by the Realm world (RMM and Realm VMs) and the Root world, so the Hypervisor in the Normal world is blocked at the hardware level from any read or write to Realm memory. CCA further encrypts Realm memory transparently as data crosses the CPU cache boundary, preventing physical-access adversaries from recovering plaintext. The hardware enforcement of GPT is the cornerstone of CCA's security model and the root cause of traditional VMI's failure on this platform: the Hypervisor can access neither the memory nor the register state of a Realm VM.

The RMM~\cite{arm2024rmm} runs at R-EL2 in the Realm world and is the manager of Realm VMs as well as a core component of their TCB. It handles the full Realm lifecycle, including creation and destruction, Stage-2 page-table maintenance (translating Realm intermediate physical addresses to physical addresses), and per-vCPU execution-context management. For each vCPU, the RMM maintains a Realm Execution Context (REC) holding the complete processor state, including general-purpose registers and key system registers such as the page-table base register TTBR1\_EL1, the exception vector base VBAR\_EL1, and the stack pointer SP\_EL0. RECs reside in Realm-protected memory and are inaccessible to the Hypervisor. Whenever a Realm VM exits due to an interrupt, exception, or voluntary yield, the RMM saves the processor state to the REC; on resumption it restores the state and returns control to the VM.

The Host interacts with the RMM through the Realm Management Interface (RMI)~\cite{arm2024rmm}, a set of standardized SMC-based (Secure Monitor Call) calls. An SMC issued by the Host traps into the EL3 Monitor, which forwards the request to the RMM. During Realm-VM execution, the central interface is RMI\_REC\_ENTER. The Host writes input arguments to the enter field of a rec\_run structure and issues the SMC; the RMM acquires the granule lock on the target REC for exclusive access, restores the processor state from the REC, and transfers control to the Realm VM at R-EL1. When the Realm VM exits, the RMM saves the state back to the REC, writes the exit reason and associated information to the exit field of rec\_run, releases the granule lock, and returns to the Host via SMC. The rec\_run structure, located in shared memory between the Normal and Realm worlds, is the only data channel between the Host and the RMM. Notably, the current RMI specification defines only interfaces required for Realm lifecycle management, with no facilities for memory introspection or runtime security monitoring of Realm VMs.

\subsection{VMI and Confidential VMs}

Virtual Machine Introspection (VMI) was first proposed by Garfinkel and Rosenblum in 2003~\cite{garfinkel2003virtual}. Its core idea is to inspect a VM's internal state, including memory contents and register values, from outside the VM in order to detect kernel-level attacks. Because the introspection runs outside the monitored VM, it remains invisible to in-VM attackers even when they have full kernel privileges.

Subsequent VMI research centered on the semantic gap problem: how to reconstruct high-level OS semantics from low-level byte-level state. Virtuoso~\cite{dolan2011virtuoso} addresses this through automated training, while VMST~\cite{fu2012space} performs online reconstruction by redirecting kernel data accesses. On the engineering side, LibVMI~\cite{payne2012simplifying} has become the de facto open-source VMI framework adopted by a wide range of subsequent work. All of these approaches, however, rely on a trusted Hypervisor to read VM memory and registers. Under the CVM threat model, where the Hypervisor is excluded from the TCB and barred by hardware from accessing VM state, this entire line of work breaks down.

To restore VMI in the CVM setting, 00SEVen~\cite{schwarz202400seven} pioneered an in-VM agent design on AMD SEV-SNP. SEV-SNP introduces VM Privilege Levels (VMPLs), allowing multiple hardware-isolated privilege tiers within a single VM. 00SEVen places a VMI agent at VMPL0, the highest tier; although the agent runs inside the VM, VMPL isolation prevents the in-VM kernel from accessing or tampering with its code or data. A remote client communicates with the VMPL0 agent via virtio-vsock, and the agent uses LibVMI together with a supplied kernel symbol table to perform checks such as process-list traversal and syscall-table integrity verification.

00SEVen's design relies on SEV-SNP-specific mechanisms (notably VMPL layering) that have no counterpart in Arm CCA, and as discussed in Section~\ref{sec:intro}, neither 00SEVen's nor traditional VMI approaches port to CCA. Realm VMs on Arm CCA therefore have no existing introspection solution.
\section{Threat Model and Design Goals}

\begin{figure}[t]
  \centering
  \includegraphics[width=\columnwidth]{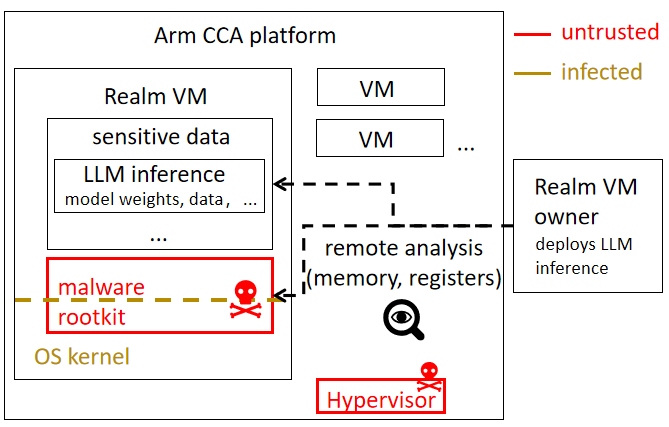}
  \caption{Application scenario: confidential LLM inference on Arm CCA.}
  \label{fig:scenario}
\end{figure}

\subsection{Application Scenario}
\label{sec:app-scenario}

LLM inference in the cloud has made privacy protection a pressing concern across both academia and industry~\cite{he2025artificial}, with two classes of sensitive assets at stake: model weights that represent substantial training investment, and user inputs that may carry personal or commercial secrets. Purely cryptographic approaches still fall short for online serving---the Multi-Party Computation (MPC)-based Transformer framework BumbleBee~\cite{jie2025bumblebee}, for instance, runs several hundred times slower than plaintext execution on large models. Trusted Execution Environments (TEEs) instead offer near-native performance with hardware-enforced isolation and are now widely viewed as the most practical option~\cite{he2025artificial}, supporting use cases that range from protecting DNN model weights against theft~\cite{li2025teeslice} to accelerating inference on TEE--GPU heterogeneous platforms~\cite{bai2025phantom} and decoupling specialized knowledge from backbone models in LLM serving~\cite{cai2025trustworthy}. Confidential VMs have emerged as a particularly compelling form of TEE for cloud AI: their VM-granularity isolation can host an entire LLM stack---OS, deep-learning framework, and model runtime---without invasive changes to the application~\cite{tan2025pipellm}.

CCA's hardware isolation, however, only shields the Realm from the outside; the VM itself runs a full OS and network services. An attacker who breaks in through a network-facing vulnerability or a supply-chain attack can deploy a rootkit to exfiltrate model weights, tamper with inference results, and hide its own activity by hooking the syscall table. As illustrated in Figure~\ref{fig:scenario}, the GPT denies the Hypervisor any access to Realm memory, so traditional Hypervisor-based VMI is structurally unable to detect such attacks.

\subsection{Threat Model}
\label{sec:threat-model}
In the traditional virtualization setting, the threat model for VMI assumes a trusted Hypervisor and considers only in-VM adversaries~\cite{garfinkel2003virtual}. An attacker exploits a network vulnerability or supply-chain attack to gain kernel control inside the VM and deploys a rootkit to hide malicious activity. VMI leverages the privileged position of the trusted Hypervisor to read VM memory and registers from outside, detecting hidden processes and tampered kernel data structures. The Hypervisor is therefore both the executor of VMI and the guarantor of VM security.

In the CVM setting, this trust assumption no longer holds. CCA explicitly excludes the Hypervisor from the Realm VM's TCB, and the GPT denies it any access to Realm memory at the hardware level, leaving the Hypervisor unable to perform VMI. The threats inside the VM, however, do not disappear: the VM still runs a full operating system and network services, and an attacker can still gain kernel control through network or supply-chain attacks. RealmEye therefore faces a stricter threat model than traditional VMI: it trusts neither the Hypervisor nor the Realm VM's operating system.

Our trust model follows Arm CCA's security architecture. The CPU hardware, the EL3 Monitor (TF-A), and the RMM are trusted and together form the Realm VM's TCB. The GPT is enforced by hardware and prevents the Normal world from reaching Realm memory. The Hypervisor and the rest of the Host software stack are untrusted: they may be controlled by a malicious cloud provider or compromised by an attacker, but CCA's hardware isolation ensures that the Hypervisor cannot read or tamper with Realm VM memory or registers. The VM owner issues VMI requests and receives results through a remote client; the client itself is trusted, but its network channel to the RMM is not, so requests and results must be protected with end-to-end encryption. The Realm VM is verified at deployment time via CCA attestation but may subsequently be compromised over the network. The RMM's integrity is upheld by hardware isolation and a secure boot chain; attacks against the RMM itself are out of scope.

A distinctive threat in the CVM setting is collusion between the untrusted Hypervisor and an in-VM attacker. If the Hypervisor can observe when a VMI scan is about to occur, it can signal the attacker (e.g., through shared memory or interrupt injection) to restore a benign kernel state in advance, presenting a clean snapshot to the scanner. To address this, RealmEye supports two trigger modes. In on-demand mode, the VM owner issues VMI requests that the Hypervisor forwards to the RMM, so the Hypervisor knows when each scan happens and collusion is possible. In periodic mode, the RMM autonomously decides scan timings at internally chosen intervals, with no dependence on Host-supplied input; from the Hypervisor's vantage point every \texttt{RMI\_REC\_ENTER} looks identical, leaving it no way to know whether a given call carries a VMI scan and effectively defeating collusion.

We focus on detecting kernel-level software attacks inside Realm VMs, and do not address hardware-level or microarchitectural threats. Physical attacks such as cold-boot attacks~\cite{halderman2009lest} are out of scope and are already mitigated by CCA's hardware memory encryption; side-channel attacks such as speculative execution~\cite{kocher2020spectre} are an orthogonal research direction complementary to this work. As for availability, a malicious Hypervisor may refuse to schedule a Realm VM or to forward VMI requests, but CCA's hardware isolation ensures it still cannot access Realm memory, so such denial-of-service behavior does not affect the correctness or confidentiality of VMI results.

\subsection{Design Goals}
\label{sec:design-goals}
RealmEye must provide full VMI capability on Arm CCA. We organize the design goals into functional and security categories. The functional goals capture the baseline VMI capabilities the system must support:

\begin{itemize}
    \item \textbf{F1 (State access):} Allow the VM owner to read memory contents and register state of a Realm VM.
    \item \textbf{F2 (Suspension):} Pause the Realm VM during a VMI scan to obtain a consistent state snapshot.
    \item \textbf{F3 (Page-level trapping):} Monitor read/write accesses to specific memory pages, capturing live modifications to kernel data structures.
    \item \textbf{F4 (API compatibility):} Expose a LibVMI-compatible interface for integration with existing VMI tooling.
\end{itemize}

CCA's threat model also imposes security requirements that traditional VMI never had to face: an untrusted Hypervisor, a potentially compromised Realm VM kernel, and an untrusted network channel can all be present simultaneously, and RealmEye must preserve both the correctness and confidentiality of VMI under all of them. Following the threat actors identified in Section~\ref{sec:threat-model}, we set the following five security goals:

\begin{itemize}
    \item \textbf{S1 (Hypervisor independence):} Neither the execution nor the correctness of VMI results depends on the Hypervisor.
    \item \textbf{S2 (VM OS independence):} Neither the execution nor the correctness of VMI results depends on the Realm VM's operating system; VMI must produce trustworthy results even when the Realm kernel is fully controlled by a rootkit.
    \item \textbf{S3 (Secure communication):} VMI results remain confidential and tamper-evident along the entire path from the RMM to the remote VM owner; the Hypervisor, the Host software stack, and intermediate network nodes see only ciphertext, and any modification is detectable.
    \item \textbf{S4 (Monitor--target separation):} The VMI logic residdes in a hardware-isolated execution domain different from the monitored VM, and no software inside the VM can access the VMI code or state.
    \item \textbf{S5 (Autonomous symbol resolution):} Locate kernel data structures under Kernel Address Space Layout Randomization (KASLR) using only RMM-controlled inputs, without relying on externally supplied symbol tables. 
\end{itemize}

S2 and S4 address two distinct concerns. S2 is about trust source: since the Realm VM may host a rootkit, VMI must read kernel memory and hardware registers directly rather than going through interfaces provided by the in-VM OS. S4 is about deployment location: the VMI logic itself must reside outside the monitored VM, fully decoupled from it. 00SEVen satisfies S2 on SEV-SNP but its VMPL0 agent still resides inside the monitored VM and relies on VMPL layering for isolation from the VM kernel, which does not meet S4. RealmEye places the VMI logic in the RMM, in an execution domain entirely separate from the Realm VM, achieving stricter isolation.
\section{Design of RealmEye}
\label{sec:design}

\begin{figure}[t]
  \centering
  \includegraphics[width=\columnwidth]{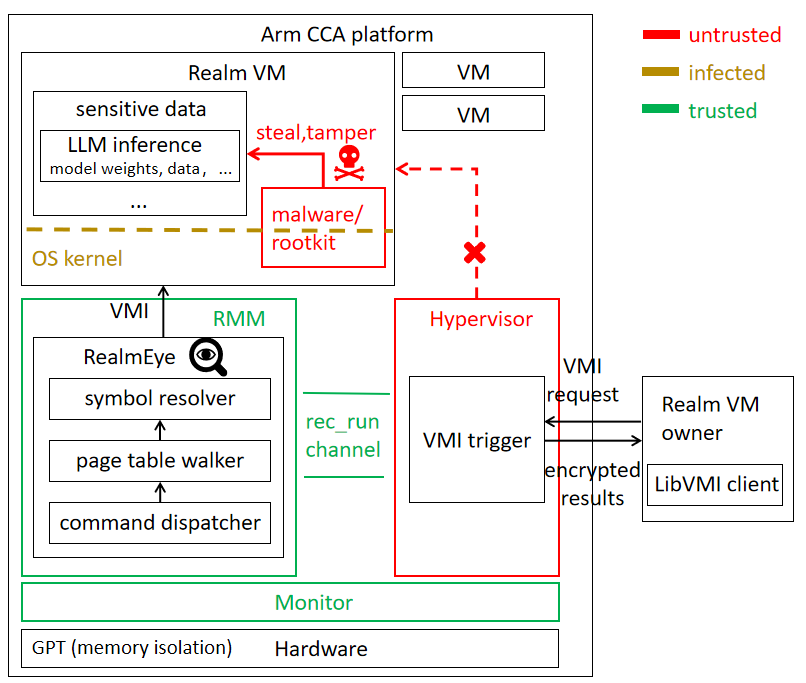}
  \caption{RealmEye system architecture.}
  \label{fig:architecture}
\end{figure}

\subsection{Design Overview}
\label{sec:design-overview}
To detect kernel-level attacks, RealmEye must let the VM owner read the memory and register state of a Realm VM, which requires its VMI logic to be hosted in a component that(i) has the hardware privilege to access Realm-internal state, and (ii) lies beyond the reach of any in-VM attacker. Few components in the Arm CCA architecture meet both conditions. The Hypervisor runs in the Normal world, is excluded from the Realm VM's TCB, and is barred by the GPT from Realm memory; it therefore cannot host VMI logic. The Realm VM itself lacks any AMD SEV-SNP VMPL-style intra-VM privilege layering, leaving no protected tier in which a VMI agent could be safely deployed. The EL3 Monitor (TF-A) is trusted but designed solely for world switching: it does not hold the Realm VM's execution context, and any VMI implementation built on top of it would still need to extend the RMM to access Realm registers and memory. The RMM is therefore the only viable choice: it runs at R-EL2 in the Realm world, is already part of the Realm VM's TCB, and has the hardware privilege to read both Realm memory and registers. Embedding VMI logic in the RMM introduces no new trust assumption and adds no code inside the Realm.

As shown in Figure~\ref{fig:architecture}, RealmEye's end-to-end VMI pipeline spans three components and faces an immediate design constraint: the RMI specification reserves no dedicated channel for VMI between the Host and the RMM. Rather than introducing a new interface, we repurpose the rec\_run shared-memory structure already used by RMI\_REC\_ENTER as the communication carrier. This structure exposes two opposing fields---enter and exit--which respectively carry Host-to-RMM requests and RMM-to-Host results; we refer to them collectively as the rec\_run channel. The two directions are used asymmetrically: the enter field embeds trigger flags and command parameters (\S\ref{sec:host-rmm-comm}), while the exit field carries encrypted scan results in an extended subfield (\S\ref{sec:host-rmm-comm}, \S\ref{sec:delivery}).

Building on this channel, the end-to-end pipeline proceeds as follows. On the VM-owner side, a LibVMI client issues a VMI request, which is forwarded by the VMI trigger module in the Hypervisor and delivered to the RMM through the enter direction of the rec\_run channel. Inside the RMM, the RealmEye engine routes the request through a command dispatcher to the appropriate handler: the page-table walker translates Realm VM virtual addresses to physical addresses and reads memory, while the symbol resolver autonomously locates kernel data structures under KASLR. Once the scan completes, the result is encrypted and returned through the exit direction, then forwarded by the Hypervisor back to the VM owner. Throughout the pipeline, the Hypervisor acts only as a relay, seeing only ciphertext and opaque flags, and learns nothing about either the request or the result.

The colors in Figure~\ref{fig:architecture} indicate three trust tiers. The RMM together with the underlying Monitor and hardware (green) is trusted and forms the Realm VM's TCB; all of RealmEye's VMI logic runs in this trusted domain. The Hypervisor (red) is untrusted: the GPT denies it any access to Realm memory at the hardware level, and its role is restricted to relaying VMI requests and ciphertext. The OS kernel (yellow) is protected by the Realm boundary but may itself be compromised over the network and host an in-kernel rootkit. Because RealmEye reads kernel data structures directly at the RMM level, bypassing every interface the in-VM OS exposes, syscall hooks installed by such a rootkit cannot influence detection results.

\subsection{RMM-Internal VMI Execution}
\label{sec:rmm-vmi}
Section~\ref{sec:design-overview} established that VMI logic should reside in the RMM. This section discusses how the four functional goals F1--F4 stated in \S\ref{sec:design-goals} are realized inside the RMM. Each goal raises a distinct challenge under the CCA architecture: F1 and F2 are constrained by the absence of any memory-introspection interface in the RMM specification, F3 must contend with the CCA-specific Stage-2 TLB coherence problem, and F4 requires interface adaptation without modifying existing VMI tools.

\textbf{F1 and F2: state access and consistent snapshots.} F1 requires reading the memory and register state of a Realm VM from the outside, while F2 requires a consistent snapshot for the duration of a scan. Register access is straightforward: the RMM reads directly from the REC, which is maintained by CPU hardware and already holds the key system registers needed for introspection---TTBR1\_EL1, VBAR\_EL1, SP\_EL0, and others---so no additional mechanism is required. Memory access, however, is constrained by the RMM specification itself, which exposes neither a Realm-VA read primitive nor an IPA read helper. Our design therefore builds a complete two-stage read path inside the RMM: starting from TTBR1\_EL1, we manually walk the Realm VM's kernel page tables to perform Stage-1 translation, invoke the existing RMM helper realm\_ipa\_to\_pa for Stage-2 translation to obtain the physical address, and finally bring the target physical page into the RMM's own address space through a transient mapping to complete the read. F2 requires no separate machinery: because every VMI scan executes inside RMI\_REC\_ENTER, before the Realm VM is allowed to resume, the VM is naturally frozen for the entire scan, and any state the RMM observes is consistent by construction. This pair of primitives forms the common substrate on which all higher-level detection strategies are built; \S\ref{sec:execution} uses it to implement process-list traversal and syscall-table integrity checking.

\textbf{F3: page-level trapping and TLB coherence.} F3 requires trapping writes to specific memory pages so that real-time modifications of kernel data structures can be detected. This is essential for catching rootkits that tamper with critical structures such as sys\_call\_table at runtime. The basic mechanism is to clear the write bit in the corresponding Stage-2 page-table entry of the Realm VM: any subsequent write triggers a Stage-2 fault that exits to the RMM. The CCA-specific challenge is TLB coherence. Stage-2 entries can be cached in the TLB, so editing the page-table contents alone does not make the permission change visible---stale entries continue to admit writes. In conventional virtualization, the Hypervisor could broadcast a cross-core TLB invalidation, but under CCA the Hypervisor is excluded from the Realm TCB and has no authority to invalidate Realm Stage-2 TLBs. The concrete design and its interaction with the existing RMM synchronization primitives are deferred to \S\ref{sec:impl}.

\textbf{F4: LibVMI-compatible interface.} F4 requires RealmEye to expose an interface compatible with LibVMI so that existing VMI tools such as DRAKVUF~\cite{lengyel2014scalability} and Volatility~\cite{volatility} can be reused without modification. We add a CCA driver backend to LibVMI that maps its core API (vmi\_read\_pa, vmi\_read\_va, vmi\_get\_vcpureg, vmi\_pause\_vm, etc.) to fine-grained VMI commands in RealmEye. Each LibVMI call enters the Host kernel via an ioctl, traverses the Host--RMM communication channel described in \S\ref{sec:host-rmm-comm}, and ultimately invokes the same primitives as F1. This design plugs RealmEye directly into the LibVMI tool ecosystem, and the implementation details follow in \S\ref{sec:libvmi}.

\subsection{Host--RMM Communication}
\label{sec:host-rmm-comm}
Section~\ref{sec:rmm-vmi} described how VMI is executed inside the RMM. This section turns to the communication path: how a VMI request travels from the Host to the RMM, and how the scan result flows back. The design of this path must address two constraints. First, the existing RMI interface is designed exclusively for Realm lifecycle management and reserves no channel for VMI. Second, the only data path available between the Host and the RMM lies in memory shared between the Normal world and the Realm world, and is therefore fully visible to the Hypervisor. We discuss RealmEye's design choices under these constraints in three parts: the triggering mechanism, the triggering modes, and the encrypted return path.

\textbf{Triggering mechanism.} The most direct design is to introduce a new SMC call through which the Host actively issues VMI requests. This conflicts with the CCA execution model. While a Realm VM is running, the granule lock of its REC is already held by the in-flight RMI\_REC\_ENTER call; any new SMC call would have to acquire the same lock and would deadlock. A natural workaround is to kick the Realm VM out via KVM and then issue the SMC, but this also fails, because KVM cannot raise a new RMI call from within the Realm-exit handling context. RealmEye therefore reuses the existing RMI\_REC\_ENTER call and embeds the trigger signal in its rec\_run.enter parameter. Just before letting a vCPU enter the Realm, the Host-side KVM module checks for pending VMI requests and, if any exist, sets a trigger bit in the parameter. On the RMM side, RMI\_REC\_ENTER inspects this bit at entry and, when set, performs the VMI scan before resuming the Realm VM. This design adds no new SMC type, preserves the semantics of the existing RMI interface, and leaves the REC-lock concurrency model untouched, embedding VMI triggering naturally into the Realm VM's execution loop.

\textbf{Triggering modes.} On top of this mechanism, RealmEye offers two triggering modes for different threat scenarios. In the on-demand mode, the VM owner issues VMI requests through the Hypervisor. Because the Hypervisor relays every request, it observes the timing of every scan and cannot defend against the collusion threat identified in \S\ref{sec:threat-model}. The periodic mode hardens against this threat. Here the trigger decision is made entirely inside the RMM at autonomously chosen intervals, with no input drawn from rec\_run.enter. From the Hypervisor's perspective, every RMI\_REC\_ENTER call looks identical, and it cannot tell which calls will trigger a scan. Since collusion relies on the Hypervisor knowing when a scan is about to occur, the periodic mode severs this channel and prevents the rootkit from restoring a clean kernel state in advance. By contrast, 00SEVen routes its VMI triggers through a Hypervisor-relayed virtio-vsock, which exposes every trigger to the Hypervisor and admits no equivalent anti-collusion design.

\textbf{Encrypted result delivery.} Scan results return from the RMM to the Host through rec\_run.exit. This structure resides in memory shared between the Normal world and the Realm world. It is the only data channel available between the Host and the RMM, and its contents are fully readable by the Hypervisor. RealmEye therefore performs encryption inside the RMM and writes only the ciphertext into rec\_run.exit. Along the entire Host-side relay path, neither the Hypervisor nor the Host kernel ever sees plaintext. The current implementation uses a pre-shared key with the VM owner; a production deployment should derive a session key through CCA attestation. Once the Host receives the ciphertext, it forwards it to the remote VM owner. The design of this final hop falls outside the scope of this section and is discussed in \S\ref{sec:trust-channel}.

\subsection{End-to-End Trust Channel}
\label{sec:trust-channel}
Section~\ref{sec:host-rmm-comm} described how RealmEye protects VMI results from being observed by the Hypervisor on the Host--RMM relay path. End-to-end security additionally requires the VM owner to confirm that the entity it talks to is in fact a legitimate RMM running on a genuine Arm CCA platform, rather than a man-in-the-middle impersonated by the Hypervisor. This goal is naturally addressed by binding remote attestation to the secure-channel protocol, so that the session key is cryptographically tied to the attested identity of the TEE.

Combining remote attestation with TLS is by now a well-established line of work in the TEE community. Knauth et al.~\cite{knauth2018integrating} introduced RA-TLS, which binds attestation evidence to the TLS handshake and laid the basic architecture for this direction. The same idea has been adopted in mainstream open-source SGX deployments~\cite{gramine2024ratls}, has been followed up by full protocol-level designs that integrate attestation into TLS 1.3 handshake and session resumption~\cite{walther2022ratls}, and is being standardized as the IETF Attested TLS draft~\cite{fossati-tls-attestation-09}. Song et al.~\cite{song2025attest} compare two embedding strategies---intra-handshake and post-handshake attestation---and observe that the post-handshake form fits long-lived connections and periodic re-attestation, a property that aligns well with how RealmEye triggers VMI in periodic mode.

Building on this line of work, RealmEye establishes trust between the RMM and the VM owner as follows. At Realm VM deployment time, CCA attestation provides the initial proof of the RMM's identity and integrity. When a VMI session is opened, the two endpoints negotiate a session key over the attested channel, and this key replaces the pre-shared key used for the symmetric encryption of VMI results in \S\ref{sec:host-rmm-comm}. For long-running deployments, post-handshake re-attestation can be performed periodically, so that the trust state of the session remains bound to the runtime integrity of the RMM. This chain ensures that the Hypervisor can neither impersonate the RMM nor tamper with VMI results in transit.
\section{Implementation}
\label{sec:impl}

This section describes the implementation of RealmEye, organized along the end-to-end data flow of a VMI operation. We first describe how VMI requests are triggered on the Host side (§\ref{sec:trigger}), then how memory and register reads as well as detection strategies are executed inside the RMM (§\ref{sec:execution}), how scan results are securely delivered back to the VM owner (§\ref{sec:delivery}), and finally how RealmEye exposes a LibVMI-compatible interface (§\ref{sec:libvmi}).

\subsection{VMI Trigger: Host-side Extensions}
\label{sec:trigger}

As discussed in §\ref{sec:host-rmm-comm}, introducing a new SMC for VMI is precluded by REC-lock contention with the in-flight RMI\_REC\_ENTER, and the KVM-kick workaround fails because KVM cannot issue a new RMI call while handling a Realm exit. RealmEye therefore embeds the VMI trigger signal directly in the input parameters of the existing RMI\_REC\_ENTER call.

The implementation spans three layers: user space, the Host kernel, and the RMM. At the user level, the VM owner issues a VMI request through a debugfs interface exposed by KVM. The Host KVM module records the request as a pending flag on the target vCPU. When KVM later invokes kvm\_rec\_enter to resume the Realm VM, it checks this flag and, if set, raises bit~63 in the enter.flags field of rec\_run as the VMI trigger signal. On the RMM side, smc\_rec\_enter examines bit~63 at entry and, when set, enters the VMI scan path before restoring control to the Realm VM. The same rec\_run channel also carries fine-grained VMI commands (e.g., a single memory read or register query) issued through an ioctl interface, with the command type and parameters passed via the general-purpose register fields of rec\_run.enter and dispatched by command type inside the RMM. Both trigger paths share the channel and require no additional SMC calls.

The correctness of this design rests on two properties. First, the VMI scan runs inside smc\_rec\_enter, where the RMM already holds the REC granule lock and has mapped the auxiliary REC granules, giving it full access to the Realm VM's registers and memory. Second, the scan completes before the Realm VM resumes, so the VM is frozen for the entire scan window and cannot modify its own memory or registers; the state observed by VMI is therefore a consistent snapshot. Once the scan finishes, the RMM clears bit~63 and proceeds normally into rec\_run\_loop, leaving the Realm VM's execution flow externally indistinguishable from an unmodified entry.

\subsection{VMI Execution: RMM-side Extensions}
\label{sec:execution}

VMI execution rests on three primitives: memory and register reads, VM suspension, and page-level trapping. VM suspension is achieved naturally as a side effect of the trigger design in §\ref{sec:trigger}: the scan completes before the Realm VM resumes, so the VM stays frozen throughout. Page-level trapping involves modifying Stage-2 page-table permission bits and entails the CCA-specific TLB coherence challenge described in \S\ref{sec:rmm-vmi} (F3).  This section focuses on memory and register reads, which form the common substrate for all higher-level detection strategies.

The read path in the RMM is a complete data-flow pipeline. It starts with register access: the RMM reads Realm VM system registers directly from the plane\_sysregs structure stored in the REC, including the page-table base TTBR1\_EL1, the exception vector base VBAR\_EL1, and the stack pointer SP\_EL0. With TTBR1\_EL1 in hand, the RMM can walk the Realm VM's kernel page tables to translate a virtual address into a physical address. Since the RMM specification provides no virtual-address translation helper, we manually implement the full ARM64 four-level Stage-1 walk, decoding page-table descriptors at levels 0 through 3. Because the Realm VM kernel uses 2\,MB huge-page mappings for parts of its address space, the walk must also handle block descriptors at levels 1 and 2, not just page descriptors at level 3. The walk yields a Realm intermediate physical address (IPA), which is then converted to a real physical address by the existing RMM helper realm\_ipa\_to\_pa for Stage-2 translation. Finally, the RMM uses buffer\_granule\_map to bring the target physical page transiently into its own address space, reads the data, and unmaps it. The entire read path executes within the Realm world; data never traverses the Normal world or appears in plaintext to the Hypervisor.

On top of this read primitive we implement process-list traversal. In the ARM64 Linux kernel, the SP\_EL0 register in kernel mode points to the task\_struct of the currently running process. The RMM uses this value, read from the REC, as the entry point of the traversal, walks the doubly-linked tasks list inside task\_struct, and reads the pid and comm (process name) fields of each task. The list-pointer and field offsets within task\_struct are extracted statically from the Realm VM's vmlinux using pahole. Because the traversal operates directly on kernel data structures and bypasses the syscall layer entirely, rootkits that hook getdents64 or similar syscalls cannot influence the result: a hidden process is still scheduled, and its task\_struct is still linked into the global list, so the RMM observes it.

Process-list traversal exposes hidden processes but does not reveal how they are hidden. We therefore add a second detection strategy that checks the integrity of the syscall table. The central difficulty is KASLR: every Realm VM boot fully randomizes kernel symbol addresses, so the RMM cannot know in advance where sys\_call\_table resides. Our solution localizes it autonomously inside the RMM, accepting no external input. The RMM first reads the hardware register VBAR\_EL1 from the REC; this register holds the base address of the exception vector table, is maintained by the CPU itself, and cannot be forged from inside the Realm VM. The ARM64 Linux kernel places the exception vector table at a fixed compile-time offset of 0x800 from the kernel text base \_stext, an offset unaffected by KASLR, so the RMM derives \_stext directly from VBAR\_EL1. The RMM then scans a bounded address range past \_stext byte-by-byte (in 8-byte units), looking for runs of eight consecutive function pointers that all fall within the kernel code segment. Because sys\_call\_table is a contiguous array of several hundred kernel function pointers, this dense pattern is essentially unique within the kernel data segment, making the false-positive probability negligible. Once located, the RMM reads the function pointer at the target syscall index; any pointer falling outside the kernel code range indicates that the syscall has been hooked.

\subsection{Result Delivery: Encrypted Communication}
\label{sec:delivery}

As established in §\ref{sec:host-rmm-comm}, scan results return through the exit field of rec\_run, which lies in shared memory readable by the Hypervisor. We therefore encrypt the result with AES-128 inside the RMM before writing it to rec\_run.exit. Our AES-128 implementation is embedded directly into the RMM runtime as roughly 100 lines of plain C with no external dependencies. The Host relays the ciphertext to the remote VM owner, who decrypts it with the pre-shared key to recover the plaintext result. The current prototype uses a hard-coded pre-shared key; a production deployment should derive a session key through CCA attestation, as discussed in §\ref{sec:trust-channel}.

Concretely, the RMM encodes each scan result as a 16-byte fixed-length plaintext, consisting of a process count (8 bytes) and a detection status code (8 bytes), and writes the AES-128 ciphertext into an extension subfield of rec\_run.exit. On Realm-VM exit, the Host KVM module reads this ciphertext and emits its hexadecimal form to the kernel log; a companion tool on the VM-owner side recovers the plaintext and extracts the process count and detection status. Fine-grained VMI commands (e.g., single-register reads) return their values through the general-purpose register fields of rec\_run.exit via the same path. As a result, all VMI results, both bulk scans and fine-grained queries, reuse the existing rec\_run channel without requiring any additional RMM--Host communication mechanism.

\subsection{LibVMI-Compatible Interface}
\label{sec:libvmi}

LibVMI provides a unified API abstraction over virtualization backends such as Xen, KVM, and Bareflank, and underpins security analysis tools including DRAKVUF and Volatility, but it has no Arm CCA backend. To plug RealmEye into this existing ecosystem, we add a CCA driver backend to LibVMI. Through this adapter, security analysts can introspect Realm VMs using the standard LibVMI API without any knowledge of CCA or the RMM internals.

The CCA driver implements the core LibVMI API, including physical-address reads (vmi\_read\_pa), virtual-address reads (vmi\_read\_va), register queries (vmi\_get\_vcpureg), VM suspend/resume (vmi\_pause\_vm and vmi\_resume\_vm), and memory-access monitoring (vmi\_set\_mem\_event). Each API call follows the same path into the RMM: LibVMI issues a KVM\_ARM\_CCA\_VMI ioctl carrying the command type and parameters to the Host KVM module, which encodes them into the general-purpose register fields of rec\_run.enter and raises the trigger flag. The RMM examines the command type at the entry of smc\_rec\_enter and dispatches it to the corresponding handler, ultimately invoking the same primitives implemented in §\ref{sec:execution}. Return values flow back through rec\_run.exit to the Host, and from there to the LibVMI caller via the ioctl. The entire path reuses the trigger channel of §\ref{sec:trigger} and the result-delivery mechanism of §\ref{sec:delivery}, introducing no new communication route.

We validated the driver on FVP through an end-to-end test: a GET\_VCPUREG command issued via LibVMI reads the Realm VM's TTBR1\_EL1, and the returned value matches what an independent full RMM scan reports for the same register. The original full-scan paths (process-list traversal and syscall-table integrity checking) show no regression after the changes. To our knowledge, this is the first LibVMI driver for Arm CCA Realm VMs.
\section{Security Analysis}
\label{sec:security}

§\ref{sec:design-goals} stated five security goals for RealmEye. We now argue, based on the design and implementation in §\ref{sec:design} and §\ref{sec:impl}, that each of them is met.

\textbf{S1 (Hypervisor Independence).}
RealmEye's entire VMI flow executes within the Realm world, and the Hypervisor plays no role in anything that affects detection correctness. The VMI logic resides in the RMM; the GPT denies the Hypervisor any access to the RMM's code and runtime state, as well as to the Realm memory the scan reads. Results are encrypted inside the RMM before being written to rec\_run.exit, leaving the Hypervisor with only ciphertext to relay. A malicious Hypervisor can refuse to forward a VMI request or clear the trigger flag in rec\_run.enter to suppress the scan altogether, but as discussed in §\ref{sec:threat-model}, this falls under availability and is out of scope; whenever a scan does occur, the Hypervisor can neither perturb its execution path nor tamper with the state it reads or the result it returns.

\textbf{S2 (VM OS Independence).}
RealmEye reads all data directly from RMM-controlled sources without invoking any interface provided by the Realm VM's OS. Register state comes from the REC, where critical registers such as VBAR\_EL1, TTBR1\_EL1, and SP\_EL0 are maintained by CPU hardware and cannot be forged from inside the VM. Kernel memory is read through the Stage-1 walk and Stage-2 translation built into the RMM (§\ref{sec:execution}), bypassing every OS-provided syscall or filesystem interface. At the detection-strategy level, process lists are obtained by traversing the task\_struct chain rather than calling ps or /proc, both of which a rootkit can poison; sys\_call\_table is located by RMM-internal pattern scanning rather than by consulting kallsyms or any OS-exposed symbol table. Even when the Realm kernel is fully controlled by a rootkit, the attacker cannot mislead VMI by tainting OS-level interfaces, so neither the execution nor the correctness of VMI depends on the in-VM OS.

\textbf{S3 (Secure Communication).}
RealmEye performs result encryption inside the RMM; plaintext never leaves the Realm world. Before writing to rec\_run.exit, the RMM encrypts the result with a key shared only with the VM owner, so no intermediary along the transport path can obtain it. The Hypervisor and the Host kernel can only read the ciphertext from rec\_run.exit and forward it onward, and any network nodes between the Host and the VM owner likewise see only ciphertext. The VM owner recovers the plaintext locally using a companion decryption tool. Across the entire path, the Hypervisor, the Host software stack, and intermediate network nodes learn nothing about the content of the scan result.

\textbf{S4 (Monitor--Target Separation).}
RealmEye places its VMI logic entirely in the RMM, in an execution domain distinct from that of the monitored Realm VM. The RMM runs at R-EL2 in the Realm world, while the Realm VM's kernel and user space run at R-EL1 and R-EL0. Arm CCA enforces isolation between these layers in hardware: no software inside the Realm VM, not even its most privileged kernel, can read the RMM's code, stack, or runtime state. Unlike 00SEVen, whose VMPL0 agent shares a VM instance with the monitored kernel and relies on VMPL software layering for isolation, RealmEye's monitor and target do not share any VM instance, achieving stricter cross-domain isolation.

\textbf{S5 (Autonomous Symbol Resolution).}
RealmEye locates kernel data structures under KASLR using only RMM-controlled inputs. The starting point is VBAR\_EL1, a CPU-maintained register that cannot be forged from inside the Realm VM. From VBAR\_EL1 and the compile-time fixed offset (0x800), the RMM derives the kernel text base \_stext; starting from \_stext, the RMM scans the kernel data segment for the dense run of code-segment pointers that uniquely identifies sys\_call\_table (§\ref{sec:execution}). The entire localization path consumes no symbol tables or symbol addresses supplied by either the Realm VM or the Hypervisor, so even if an attacker injects fake symbol information through these channels, the data simply never enters the RMM's localization procedure.
\section{Evaluation}
\label{sec:eval}

\subsection{Experimental Setup}
\label{sec:setup}

We evaluate RealmEye on the Arm Fixed Virtual Platform (FVP). The platform is an FVP\_Base\_RevC-2xAEMvA model running Fast Models 11.29.42, configured with two clusters of four cores each (8 cores total), 4\,GB of DRAM, and Armv9.2-A support including the RME extension. 
The software stack builds on the Arm CCA reference implementation(cca-3world)~\cite{cca3world}, comprising TF-RMM v0.1.0~\cite{tfrmm}(which we extend with VMI functionality), TF-A~\cite{tfa} as the EL3 Monitor, and Linux 6.15.0-rc1 (cca-host/v8 branch)~\cite{linuxcca} as both the Host and Realm VM kernel. Realm VMs are launched through kvmtool~\cite{kvmtool}. Performance is measured using the ARM64 CNTPCT\_EL0 counter to record clock cycles. Note that FVP is configured with cache\_state\_modelled=0 (no cache behavior modeling), so the reported numbers reflect instruction-level execution cost rather than absolute latency on real hardware.

\subsection{Rootkit Detection}

We evaluate RealmEye against Diamorphine, an open-source LKM rootkit widely used on ARM64 Linux and adopted as a benchmark in prior work~\cite{schwarz202400seven, orenbach2025blueguard}. Diamorphine exhibits both process hiding and syscall-table hooking, two canonical kernel-level stealth techniques, making it a representative target. We test the two attack classes with two corresponding detection strategies: process-list traversal verifies whether RealmEye can uncover rootkit-hidden activities, while syscall-table integrity checking verifies whether it can further pinpoint the technique used.

\textbf{Process-list traversal.} The strategy bypasses rootkit-poisoned user-space interfaces and reads the kernel task\_struct list directly from the RMM, exposing processes that are no longer visible to tools like ps but are still scheduled by the kernel. We launch a test process in a Realm VM, load Diamorphine, and use its control interface to mark the process as hidden; ps and ls /proc inside the VM no longer report it. When RealmEye traverses the task\_struct list from the RMM, the hidden process's pid and comm fields appear in the result in full. As a control, on a clean kernel without Diamorphine the RMM output matches the in-VM ps exactly, with no missing or spurious entries, confirming that the detection logic itself is correct and that the discovery in the attack scenario is genuinely attributable to the rootkit. This outcome reflects an inherent limitation of Diamorphine's hiding mechanism: by hooking getdents64 to filter /proc readouts, it only masks the user-space view and cannot perturb the in-kernel task\_struct list. Hidden processes are still scheduled normally, and their metadata remains intact in the global process chain. Because RealmEye reads kernel data structures directly at the RMM level without traversing any syscall-layer interface, hiding mechanisms that rely on syscall hooks are simply ineffective against it.

\textbf{Syscall-table integrity checking.} The strategy verifies whether each entry in sys\_call\_table still points into the kernel code segment; an entry pointing elsewhere (e.g., into the module-loading region) indicates a hooked syscall. The autonomous localization of sys\_call\_table under KASLR is described in §\ref{sec:execution}. We run the check before and after loading Diamorphine, focusing on the two entries it tampers with: getdents64 and kill. On the clean kernel both entries fall within the kernel code segment and match the corresponding symbol addresses in /proc/kallsyms; after loading Diamorphine, both entries are replaced by hook-function addresses in the module-loading region, and the check correctly flags them as tampered. The sys\_call\_table base address derived autonomously by the RMM matches the kallsyms entry in both runs, confirming that the localization scheme of §\ref{sec:execution} is reliable under KASLR; we observe no false positives.

The two strategies together cover two complementary dimensions of rootkit detection. Process-list traversal exposes state divergence at the kernel data-structure level, surfacing the fact that hidden malicious activity exists; syscall-table integrity checking pinpoints the technical vehicle of the attack at the kernel code-flow level, revealing how the hiding is implemented. Both checks run entirely inside the RMM, depend on no Realm-VM interfaces, and accept no Hypervisor-supplied input, in line with the design goals of §\ref{sec:design-goals}.

\subsection{Microbenchmarks}
\label{sec:eval-micro}
\begin{figure}[t]
  \centering
  \includegraphics[width=\columnwidth]{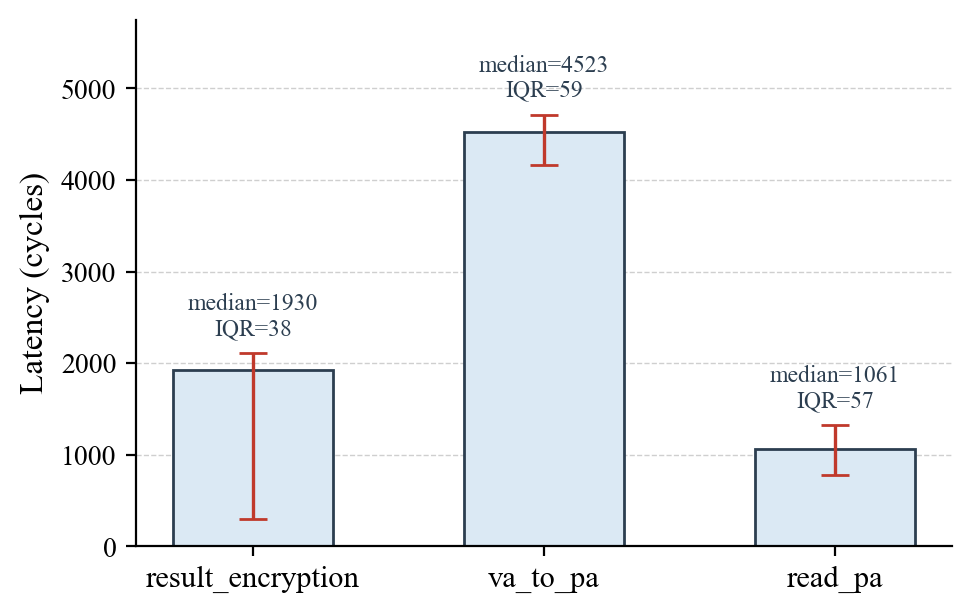}
  \caption{Microbenchmark latency (CPU cycles, $n=1000$) of the three RMM-internal primitives. Error bars span min to max; medians and IQRs are labeled on top of each bar.}
  \label{fig:microbench}
\end{figure}

To understand the cost composition of RealmEye's end-to-end VMI operations, we first measure the core primitives that the RMM invokes repeatedly. As described in §\ref{sec:impl}, all higher-level detection strategies ultimately reduce to three RMM-internal primitives: result\_encryption (encrypting the VMI output before delivery), vmi\_va\_to\_pa (translating a Realm kernel virtual address to a physical address), and vmi\_read\_pa (reading 8 bytes of Realm physical memory inside the RMM). The three correspond, respectively, to the encrypted-delivery component of §\ref{sec:delivery}, the four-level page-table walk of §\ref{sec:execution}, and the transient page-mapping read of §\ref{sec:execution}.

We measure each primitive inside the RMM using the difference between two mrs CNTPCT\_EL0 reads taken before and after invocation. Each primitive is executed 1010 times; we discard the first 10 warmup samples and retain 1000. Under FVP simulation, the resolution of CNTPCT\_EL0 occasionally produces samples that fall abnormally low into the 0--300 cycles range, giving all three primitives a distribution of "tightly concentrated majority with a few extremely low outliers." To prevent these outliers from skewing the report, we use the median (p50) and interquartile range (IQR) as the primary metrics, while showing the full min--max range as error bars for transparency. As noted in §\ref{sec:setup}, FVP runs with cache\_state\_modelled=0 and models neither L1/L2 caches nor TLBs, so the numbers in this section reflect instruction-level relative cost rather than absolute latency on real hardware.

Figure~\ref{fig:microbench} summarizes the three primitives. All medians fall within a few thousand cycles: vmi\_va\_to\_pa is the heaviest at 4523 cycles, result\_encryption sits in the middle at 1930 cycles, and vmi\_read\_pa is the lightest at 1061 cycles. The IQR of all three stays within 60 cycles, an extremely tight spread that confirms the strong determinism of the in-RMM VMI execution path. These costs are negligible compared to a full SMC world-switch or KVM round-trip (§\ref{sec:eval-macro} provides the comparison data), meaning that the end-to-end latency of higher-level VMI operations is dominated by the number of primitive invocations rather than by the cost of any individual call. We now analyze the three primitives in order of decreasing cost.

The cost composition of vmi\_va\_to\_pa can be verified independently from the data: a four-level walk dominated by four underlying physical-memory reads should cost approximately $4 \times 1061 = 4244$ cycles, and the measured median of 4523 cycles deviates from this estimate by only about 280 cycles (roughly 6\%), accounting for the function call frame, per-level page-table index computation, and block-descriptor checking for huge-page mappings. We note that despite its name, vmi\_read\_pa takes a Realm IPA rather than a real physical address as input; the Stage-2 translation through realm\_ipa\_to\_pa is already included in the 1061 cycles, and the naming follows the semantics from the VMI caller's perspective.

In summary, the three primitives form a clean and predictable cost structure: vmi\_read\_pa is the smallest unit of memory-access cost (1061 cycles); vmi\_va\_to\_pa is built linearly from four vmi\_read\_pa calls (4523 cycles, within 6\% of the $4\times$ estimate); and result\_encryption is a one-time per-command fixed cost with negligible impact (1930 cycles). This structure implies that the total latency of higher-level VMI operations is linearly predictable from the invocation counts of these primitives, and that no individual primitive constitutes a performance bottleneck. The next section validates this prediction through three macrobenchmark experiments and reveals where end-to-end latency actually comes from.

\subsection{Macrobenchmarks}
\label{sec:eval-macro}

The microbenchmarks in \S\ref{sec:eval-micro} characterize the cost of RealmEye's three core primitives. We now evaluate two complete detection strategies end to end---process-list walking and syscall-table integrity checking---to test whether the total latency of higher-level VMI operations can be linearly predicted from primitive costs. Both strategies run entirely inside the RMM, and we sample CNTPCT\_EL0 at strategy entry and exit.

\subsubsection{Process List Walking}
\label{sec:eval-procwalk}

Starting from the task\_struct of the running process, this strategy walks the kernel task list and reads the comm and pid fields of each entry. The entry point is derived from SP\_EL0 (\S\ref{sec:rmm-vmi}); the implementation is given in \S\ref{sec:execution}. We measure on a Realm VM running 50 user-space processes, executing the strategy 55 times and retaining 50 samples after a 5-run warmup.

The measured median latency is 1{,}161{,}396 cycles, with an IQR of only 13 cycles. This result can be predicted independently from the microbenchmark data. Each task\_struct visit incurs five vmi\_va\_to\_pa calls: the upper and lower halves of comm, pid, the tasks list pointer, and the self-referential anchor of the loop. Using the median 4523 cycles from \S\ref{sec:eval-micro}, 50 processes give an estimate of $\sim$1.13\,M cycles. The estimate deviates from the measured median by about 30{,}646 cycles ($\sim$2.6\%), accounting for loop control flow and log output. The linear prediction of the microbenchmarks holds for this strategy.

\subsubsection{Syscall Table Integrity Check}
\label{sec:eval-syscall}

This strategy derives the kernel text base \_stext from VBAR\_EL1 and scans a one-megabyte range past \_stext for a contiguous array of kernel code-segment pointers matching the shape of sys\_call\_table; the localization algorithm is detailed in \S\ref{sec:execution}. Because KASLR fixes the address layout once the Realm boots, the scan path is fully deterministic inside the RMM, and a single measurement suffices.

The measured total latency is 192{,}246{,}222 cycles, with sys\_call\_table located at scan offset 0x40a08 (about 264\,KB, the first 26\% of the range). This cost is again predictable from the microbenchmarks: the scan visits 33{,}029 candidate start positions, each requiring on average $\sim$1.3 vmi\_va\_to\_pa calls, since most candidates terminate after the first pointer read once the target falls outside the kernel code range. The estimated total is $\sim$194\,M cycles, within 1\% of the measurement. The linear prediction holds for this strategy as well.

\subsubsection{Summary}
\label{sec:eval-summary}

Table~\ref{tab:macrobench} summarizes the measured and predicted latencies of the two detection strategies, along with the relative error. The prediction error stays within 3\% in both cases, confirming that the cost model from \S\ref{sec:eval-micro} carries over to complete detection strategies: the latency of higher-level VMI operations is linearly predictable from the invocation counts of the underlying primitives.

\begin{table*}[!t]
\centering
\caption{Detection latency: measured vs.\ predicted. (\S\ref{sec:eval-micro}).}
\label{tab:macrobench}
\small
\begin{tabular}{@{}lrrr@{}}
\toprule
\textbf{Detection Strategy} & \textbf{Measured} & \textbf{Predicted} & \textbf{Error} \\
                            & \textbf{(cycles)} & \textbf{(cycles)}  &                \\
\midrule
Process List Walking (50 procs) & 1{,}161{,}396   & 1{,}130{,}750         & 2.6\% \\
Syscall Table Integrity Check   & 192{,}246{,}222 & $\sim$194{,}000{,}000 & $\sim$1\% \\
\bottomrule
\end{tabular}
\end{table*}

\subsection{Per-Command Framework Cost}
\label{sec:eval-framework}

Sections~\ref{sec:eval-micro} and~\ref{sec:eval-macro} characterize how VMI cost scales with detection complexity, but leave aside the framework's fixed communication overhead---the toll that every VMI command must pay regardless of its workload. We characterize this cost using the lightest fine-grained command, \mbox{GET\_VCPUREG}.

GET\_VCPUREG simply reads a saved system register (e.g., TTBR1\_EL1) from the REC, with no address translation or memory-read primitive involved. Its end-to-end latency therefore approximates the framework's fixed overhead, covering the user-space ioctl, the vCPU kick and wait\_event blocking inside KVM, the SMC world switch, dispatch inside the RMM, and the reverse path through rec\_run and handle\_rec\_exit. We wrap the ioctl with clock\_gettime(CLOCK\_MONOTONIC) on the Host and run 1000 invocations.

The measured median latency is 502\,\textmu s, with an IQR of about 33\,\textmu s (6.7\% of the median). For comparison, the heaviest in-RMM primitive in \S\ref{sec:eval-micro} costs only 4523 cycles, nowhere near enough to account for 502\,\textmu s of end-to-end latency. The bulk of a single VMI command is therefore spent outside the RMM, on the SMC world switch and the KVM path.

This result splits RealmEye's overhead into two parts. The cost of higher-level detection logic accumulates from primitive invocations and can be reduced through more efficient detection algorithms. The framework's fixed overhead, by contrast, stems from CCA's existing Host--RMM communication path and imposes a $\sim$502\,\textmu s floor on every VMI command; pushing this floor down further requires batched commands or extensions to the RMI interface, which we leave to future work.
\section{Discussion}
\label{sec:discussion}

\subsection{OpenCCA and Future Performance Evaluation}
\label{sec:disc-opencca}

All performance numbers in \S\ref{sec:eval} come from the Arm FVP, configured with cache\_state\_modelled=0. FVP models neither L1/L2 caches, TLBs, nor the pipeline, so the reported cycles reflect instruction-level relative cost rather than absolute latency on real hardware. This limitation has concrete effects on several measurements: the 4523 cycles of vmi\_va\_to\_pa in \S\ref{sec:eval-micro}, for instance, come from four sequential physical-memory reads whose per-access latency on real hardware would vary substantially with TLB and cache behavior, and the 502\,\textmu s end-to-end cost of GET\_VCPUREG in \S\ref{sec:eval-framework} is dominated by SMC world switches whose real-hardware cost likewise requires re-measurement. More importantly, the FVP limitation prevents us from quantifying the benefit of the optimizations proposed at the end of \S\ref{sec:eval}, such as batched VMI commands.

OpenCCA~\cite{bertschi2025opencca} provides a viable path to real-hardware evaluation. By modifying TF-A, the RMM, and U-Boot, OpenCCA ports the CCA execution model to commodity Armv8.2 boards without RME (e.g., Radxa Rock 5b), while leaving the Hypervisor, CVM kernel, and VMM untouched. RealmEye's changes are concentrated in the RMM and the Host KVM module, orthogonal to the firmware components OpenCCA modifies, so it should port to OpenCCA in a lift-and-shift manner. A real-hardware run on a board such as the RK3588 would clarify the absolute latencies of our microbenchmarks, the actual split between SMC switches and the KVM path in \S\ref{sec:eval-framework}, and the Stage-2 TLB-invalidation cost central to F3 in \S\ref{sec:rmm-vmi}---all of which FVP cannot capture.

OpenCCA does not, by design, offer security guarantees equivalent to genuine CCA: it emulates RME world switching and GPT configuration in software for the purposes of performance evaluation and functional validation, not secure deployment. Running RealmEye on OpenCCA would therefore serve only to obtain more accurate performance estimates; its security argument (\S\ref{sec:security}) continues to rest on the functional validation under FVP and the hardware guarantees of CCA itself. As no commercial CCA hardware is yet available, we leave a real-hardware port of RealmEye to OpenCCA as future work.

\subsection{VMI on Other CVM Platforms}
\label{sec:disc-cvm}

Confidential VMs are not unique to Arm CCA: AMD SEV-SNP and Intel TDX are the two other mainstream platforms, and both face the same breakdown of traditional VMI since they too exclude the Hypervisor from the VM's TCB. SEV-SNP already has a direct counterpart to RealmEye in 00SEVen (\S\ref{sec:disc-sev}); TDX has no equivalent and remains an open research question (\S\ref{sec:disc-tdx}).

\subsubsection{SEV-SNP and 00SEVen}
\label{sec:disc-sev}
\begin{table*}[t]
\centering
\caption{Comparison between 00SEVen and RealmEye.}
\label{tab:00seven-vs-realmeye}
\small
\begin{tabular}{@{}lll@{}}
\toprule
\textbf{Aspect} & \textbf{00SEVen} & \textbf{RealmEye} \\
\midrule
Platform                  & AMD SEV-SNP                          & Arm CCA                          \\
VMI deployment            & VMPL0 (in-VM)                        & RMM (out-of-VM)                  \\
Monitor--target separation & VMPL software layering              & Cross-domain hardware isolation  \\
Trigger channel           & virtio-vsock (visible to Hypervisor) & rec\_run (opaque to Hypervisor)  \\
Anti-collusion            & Not supported                        & Periodic mode                    \\
Symbol resolution         & External System.map                  & Autonomous (VBAR\_EL1 + scanning) \\
\bottomrule
\end{tabular}
\end{table*}

00SEVen is the first CVM-VMI system on SEV-SNP. It exploits VMPL to host a hardware-protected VMI agent at VMPL0, the highest in-VM privilege tier, shielded from the VM kernel running at VMPL3. A remote client communicates with the agent over virtio-vsock; the agent uses LibVMI together with an externally supplied System.map to resolve kernel symbols and run checks such as process-list traversal and syscall-table integrity verification.

The deployment-location choice in each system is dictated by its underlying isolation architecture. SEV-SNP's VMPL provides multiple in-VM privilege tiers, so 00SEVen can safely deploy its agent inside the VM. Arm CCA offers no such intra-VM layering, and the RMM outside the VM is the only trusted component available to RealmEye. This constraint, however, yields stricter monitor--target separation (S4): 00SEVen's agent shares a VM instance with the monitored kernel and relies on VMPL software layering for isolation, whereas RealmEye's VMI logic sits in the RMM, in a fully separate execution domain enforced by CCA hardware. The difference reflects platform constraints rather than a critique of 00SEVen's design.

RealmEye additionally defends against two threats that 00SEVen does not. First, collusion: every 00SEVen trigger goes through a Hypervisor-relayed virtio-vsock, exposing scan timing to a Hypervisor that may collude with an in-VM rootkit to restore a clean state in advance. RealmEye's periodic mode places the trigger decision entirely inside the RMM, hiding scan timing from the Hypervisor and severing this channel (\S\ref{sec:host-rmm-comm}). Second, symbol input: 00SEVen relies on an externally supplied System.map, which under RealmEye's threat model would have to traverse the untrusted Hypervisor and could be tampered with; RealmEye localizes symbols autonomously from VBAR\_EL1 and pattern scanning (\S\ref{sec:execution}). Note that 00SEVen's threat model assumes neither Hypervisor--rootkit collusion nor tampered symbol input, so these are not criticisms of its design but RealmEye's responses to the stronger threat model that CCA imposes. Table~\ref{tab:00seven-vs-realmeye} summarizes the differences.

\subsubsection{TDX}
\label{sec:disc-tdx}

Intel TDX, the third mainstream CVM platform, runs a TDX module in SEAM mode that plays the same role as the RMM: a trusted intermediary for the Trust Domain (the TDX equivalent of a Realm VM), more privileged than the TD and independent of the untrusted Hypervisor. RealmEye's core principles---in-module VMI execution, encrypted return, autonomous symbol resolution---apply to TDX in concept. The TDX module's extension mechanism, SEAMCALL semantics, and attestation protocol differ from CCA, however, and porting RealmEye to TDX is a separate research path. To our knowledge, no public TDX-VMI work exists.

\section{Related Work}
\label{sec:related}

\subsection{Traditional VMI}
\label{sec:related-vmi}

VMI was first proposed by Garfinkel and Rosenblum in 2003~\cite{garfinkel2003virtual}: read VM memory and registers from outside via a trusted Hypervisor to detect kernel-level attacks. Two decades of follow-up work have shaped the field along several lines. The first is the semantic gap, the central challenge of reconstructing high-level OS semantics from low-level byte state. Jiang et al.~\cite{jiang2007stealthy} introduced guest view casting in VMwatcher to detect stealthy malware by comparing in-VM and out-of-VM views; Virtuoso~\cite{dolan2011virtuoso} narrows the gap through automated training, while VMST~\cite{fu2012space} performs online reconstruction by redirecting kernel data accesses. A second line extends VMI from passive observation to active monitoring: Lares~\cite{payne2008lares} places hooks inside the guest while keeping the analysis in an isolated VM, and Process Out-Grafting~\cite{srinivasan2011process} relocates a target process out of the monitored VM for fine-grained execution monitoring. On the engineering side, LibVMI~\cite{payne2012simplifying} provides a unified introspection API across virtualization backends, and DRAKVUF~\cite{lengyel2014scalability} builds a large-scale dynamic malware analysis framework on top of it. More recent work targets performance, e.g., offloading introspection to DPUs~\cite{orenbach2025blueguard}.

All of these, however, presuppose a trusted Hypervisor as both the executor of VMI and the only privileged accessor of VM state. This assumption no longer holds in CVMs (\S\ref{sec:threat-model}), and the traditional model breaks down. RealmEye stays compatible with LibVMI (\S\ref{sec:libvmi}) but shifts the executor from the Hypervisor to the RMM, extending this line of work to the CVM era.

\subsection{Arm CCA Security Solutions}
\label{sec:related-cca}

Since its 2021 announcement, Arm CCA has become one of the most active areas in confidential computing, with work spanning several orthogonal directions that together extend the platform's security model and capabilities. RealmEye targets a direction not addressed by any of these efforts---introspection of Realm VMs---and complements them as part of the broader CCA research landscape.

\noindent\textbf{Formal verification.} The correctness of the RMM, as CCA's root of trust, is foundational. Fox et al.~\cite{fox2023verification} verify the RMM specification and prototype using interactive theorem proving, model checking, and concurrency-aware testing, providing the first formal-verification effort for CCA firmware.

\noindent\textbf{Comparison with TrustZone.} As Arm's next-generation TEE, CCA differs substantially from the existing TrustZone in design goals and capabilities. Huang et al.~\cite{huang2024sok} systematically compare the two along the axes of flexibility, security, and performance, offering guidance for choosing between Arm TEEs in different scenarios.

\noindent\textbf{Application- and container-level isolation.} CCA targets VM-granularity confidentiality, leaving application- and container-level isolation underexplored. SHELTER~\cite{zhang2023shelter,zhang2025complementing} leverages CCA hardware primitives to build lightweight user-space application sandboxes, allowing third-party developers to deploy protected apps without trusting the host OS. RContainer~\cite{zhou2025rcontainer} extends these primitives to container granularity, using a trusted mini-OS and a GPC-based con-shim to protect containers under an untrusted OS.

\noindent\textbf{Accelerator extensions.} Real-world confidential workloads, especially AI inference, increasingly rely on accelerators that CCA does not natively cover. CAGE~\cite{wang2024cage,wang2025building} and ACAI~\cite{sridhara2024acai} extend CCA to GPUs and generic accelerators respectively: CAGE introduces a shadow-task mechanism to support unified-memory GPUs without hardware modification, while ACAI extends isolation to the device bus and IOMMU through CCA's GPC mechanism.

\noindent\textbf{Real-hardware evaluation.} With no commercial CCA hardware yet available, OpenCCA~\cite{bertschi2025opencca} ports the CCA execution model to commodity Armv8.2 boards (e.g., RK3588) by modifying TF-A, the RMM, and U-Boot, providing the first low-cost real-hardware evaluation path. As discussed in \S\ref{sec:disc-opencca}, we plan to port RealmEye to OpenCCA for real-hardware measurement in future work.

\section{Conclusion}
\label{sec:conclusion}

We presented RealmEye, the first VMI system for Arm CCA Realm VMs. By placing all VMI logic inside the RMM with no agent inside the Realm VM, RealmEye achieves hardware-enforced separation between monitor and target. We identified three CCA-specific challenges---the absence of memory-introspection interfaces in the RMM, the concurrency constraint between the REC lock and VMI triggers, and Stage-2 TLB coherence---and addressed them with a complete in-RMM read path, a non-intrusive trigger mechanism that reuses RMI\_REC\_ENTER, and a periodic mode that resists Hypervisor--rootkit collusion. The implementation spans three components of the CCA software stack and forms an end-to-end pipeline from Host triggering to encrypted result delivery. Evaluation on the Arm FVP shows that RealmEye effectively detects both process hiding and syscall-table hooking, and that the cost of in-RMM VMI operations is linearly predictable from primitive invocation counts, with end-to-end latency dominated by SMC world switches and the KVM path. Future work includes real-hardware evaluation, extension to TDX, and finer-grained kernel-data semantic validation.





\bibliographystyle{IEEEtran}
\bibliography{refs}

@article{sev2020strengthening,
  title={Strengthening VM isolation with integrity protection and more},
  author={Sev-Snp, AMD},
  journal={White Paper, January},
  volume={53},
  number={2020},
  pages={1450--1465},
  year={2020}
}

@manual{arm2023rme,
  author={{Arm Limited}},
  title={{Arm Architecture Reference Manual Supplement: The Realm Management Extension(RME),for Armv9-A}},
  organization={Arm Limited},
  year={2023},
  number={DDI 0615},
  url={https://developer.arm.com/documentation/ddi0615/latest/},
  note={Document number: DDI 0615. Accessed: 2026-05-05}
}

@inproceedings{bai2025phantom,
  title={Phantom: {Privacy-Preserving} Deep Neural Network Model Obfuscation in Heterogeneous {TEE} and {GPU} System},
  author={Bai, Juyang and Chowdhuryy, Md Hafizul Islam and Li, Jingtao and Yao, Fan and Chakrabarti, Chaitali and Fan, Deliang},
  booktitle={34th USENIX Security Symposium (USENIX Security 25)},
  pages={5565--5582},
  year={2025}
}

@inproceedings{bertschi2025opencca,
  title={OpenCCA: An open framework to enable arm cca research},
  author={Bertschi, Andrin and Shinde, Shweta},
  booktitle={2025 IEEE European Symposium on Security and Privacy Workshops (EuroS\&PW)},
  pages={429--434},
  year={2025},
  organization={IEEE}
}

@article{cai2025trustworthy,
  title={Trustworthy and Controllable Professional Knowledge Utilization in Large Language Models with TEE-GPU Execution},
  author={Cai, Yifeng and An, Zhida and Meng, Yuhan and Liu, Houqian and Wang, Pengli and Lei, Hanwen and Guo, Yao and Li, Ding},
  journal={arXiv preprint arXiv:2512.16238},
  year={2025}
}

@inproceedings{dolan2011virtuoso,
  title={Virtuoso: Narrowing the semantic gap in virtual machine introspection},
  author={Dolan-Gavitt, Brendan and Leek, Tim and Zhivich, Michael and Giffin, Jonathon and Lee, Wenke},
  booktitle={2011 IEEE symposium on security and privacy},
  pages={297--312},
  year={2011},
  organization={IEEE}
}

@techreport{fossati-tls-attestation-09,
    number =    {draft-fossati-tls-attestation-09},
    type =      {Internet-Draft},
    institution =   {Internet Engineering Task Force},
    publisher = {Internet Engineering Task Force},
    note =      {Work in Progress},
    url =       {https://datatracker.ietf.org/doc/draft-fossati-tls-attestation/09/},
    author =    {Hannes Tschofenig and Yaron Sheffer and Paul Howard and Ionuț Mihalcea and Yogesh Deshpande and Arto Niemi and Thomas Fossati},
    title =     {{Using Attestation in Transport Layer Security (TLS) and Datagram Transport Layer Security (DTLS)}},
    pagetotal = 34,
    year =      2025,
    month =     apr,
    day =       30,
}

@article{fox2023verification,
  title={A verification methodology for the arm{\textregistered} confidential computing architecture: From a secure specification to safe implementations},
  author={Fox, Anthony CJ and Stockwell, Gareth and Xiong, Shale and Becker, Hanno and Mulligan, Dominic P and Petri, Gustavo and Chong, Nathan},
  journal={Proceedings of the ACM on Programming Languages},
  volume={7},
  number={OOPSLA1},
  pages={376--405},
  year={2023},
  publisher={ACM New York, NY, USA}
}

@inproceedings{fu2012space,
  title={Space traveling across vm: Automatically bridging the semantic gap in virtual machine introspection via online kernel data redirection},
  author={Fu, Yangchun and Lin, Zhiqiang},
  booktitle={2012 IEEE symposium on security and privacy},
  pages={586--600},
  year={2012},
  organization={IEEE}
}

@inproceedings{garfinkel2003virtual,
  title={A virtual machine introspection based architecture for intrusion detection.},
  author={Garfinkel, Tal and Rosenblum, Mendel and others},
  booktitle={Ndss},
  volume={3},
  number={2003},
  pages={191--206},
  year={2003},
  organization={San Diega, CA}
}

@misc{gramine2024ratls,
  title        = {{Gramine}: A Library OS for Unmodified Applications with {RA-TLS} and Secret Provisioning},
  author       = {{Gramine Project}},
  howpublished = {Gramine documentation},
  year         = {2024},
  note         = {Available at: https://gramine.readthedocs.io/en/stable/attestation.html; accessed 2026-05-06}
}

@article{halderman2009lest,
  title={Lest we remember: cold-boot attacks on encryption keys},
  author={Halderman, J Alex and Schoen, Seth D and Heninger, Nadia and Clarkson, William and Paul, William and Calandrino, Joseph A and Feldman, Ariel J and Appelbaum, Jacob and Felten, Edward W},
  journal={Communications of the ACM},
  volume={52},
  number={5},
  pages={91--98},
  year={2009},
  publisher={ACM New York, NY, USA}
}

@inproceedings{huang2024sok,
  title={SoK: A comparison study of arm TrustZone and CCA},
  author={Huang, Haoyang and Zhang, Fengwei and Yan, Shoumeng and Wei, Tao and He, Zhengyu},
  booktitle={2024 International Symposium on Secure and Private Execution Environment Design (SEED)},
  pages={107--118},
  year={2024},
  organization={IEEE}
}

@techreport{intel2023tdx,
  author={{Intel Corporation}},
  title={{Intel\textregistered{} Trust Domain Extensions}},
  institution={Intel Corporation},
  type={White Paper},
  year={2023},
  url={https://www.intel.com/content/www/us/en/developer/tools/trust-domain-extensions/overview.html},
  note={Accessed: 2026-05-05}
}

@article{knauth2018integrating,
  title={Integrating remote attestation with transport layer security},
  author={Knauth, Thomas and Steiner, Michael and Chakrabarti, Somnath and Lei, Li and Xing, Cedric and Vij, Mona},
  journal={arXiv preprint arXiv:1801.05863},
  year={2018}
}

@article{kocher2020spectre,
  title={Spectre attacks: Exploiting speculative execution},
  author={Kocher, Paul and Horn, Jann and Fogh, Anders and Genkin, Daniel and Gruss, Daniel and Haas, Werner and Hamburg, Mike and Lipp, Moritz and Mangard, Stefan and Prescher, Thomas and others},
  journal={Communications of the ACM},
  volume={63},
  number={7},
  pages={93--101},
  year={2020},
  publisher={ACM New York, NY, USA}
}

@inproceedings{lengyel2014scalability,
  title={Scalability, fidelity and stealth in the DRAKVUF dynamic malware analysis system},
  author={Lengyel, Tamas K and Maresca, Steve and Payne, Bryan D and Webster, George D and Vogl, Sebastian and Kiayias, Aggelos},
  booktitle={Proceedings of the 30th annual computer security applications conference},
  pages={386--395},
  year={2014}
}

@article{li2025teeslice,
  title={Teeslice: Protecting sensitive neural network models in trusted execution environments when attackers have pre-trained models},
  author={Li, Ding and Zhang, Ziqi and Yao, Mengyu and Cai, Yifeng and Guo, Yao and Chen, Xiangqun},
  journal={ACM Transactions on Software Engineering and Methodology},
  volume={34},
  number={6},
  pages={1--49},
  year={2025},
  publisher={ACM New York, NY}
}

@inproceedings{li2022design,
  title={Design and verification of the arm confidential compute architecture},
  author={Li, Xupeng and Li, Xuheng and Dall, Christoffer and Gu, Ronghui and Nieh, Jason and Sait, Yousuf and Stockwell, Gareth},
  booktitle={16th USENIX Symposium on Operating Systems Design and Implementation (OSDI 22)},
  pages={465--484},
  year={2022}
}

@incollection{jie2025bumblebee,
  title={BumbleBee: Secure two-party inference framework for large transformers},
  author={jie Lu, Wen and Huang, Zhicong and Gu, Zhen and Li, Jingyu and Liu, Jian and Hong, Cheng and Ren, Kui and Wei, Tao and Chen, WenGuang},
  booktitle={NDSS 2025},
  year={2025},
  publisher={The Internet Society}
}

@inproceedings{orenbach2025blueguard,
  title={{BlueGuard}: Accelerated Host and Guest Introspection Using {DPUs}},
  author={Orenbach, Meni and Ailabouni, Rami and Masalha, Nael and Nguyen, Thanh and Saleh, Ahmad and Block, Frank and Alder, Fritz and Arkin, Ofir and Atamli, Ahmad},
  booktitle={34th USENIX Security Symposium (USENIX Security 25)},
  pages={645--664},
  year={2025}
}

@techreport{payne2012simplifying,
  title={Simplifying virtual machine introspection using LibVMI.},
  author={Payne, Bryan D},
  year={2012},
  institution={Sandia National Laboratories}
}

@inproceedings{schwarz202400seven,
  title={{00SEVen}--Re-enabling Virtual Machine Forensics: Introspecting Confidential {VMs} Using Privileged {in-VM} Agents},
  author={Schwarz, Fabian and Rossow, Christian},
  booktitle={33rd USENIX Security Symposium (USENIX Security 24)},
  pages={1651--1668},
  year={2024}
}

@inproceedings{song2025attest,
  title={When to Attest? Intra-and Post-Handshake Attestation for IoT Swarms},
  author={Song, Yuxuan and Sardar, Muhammad Usama and Fedrecheski, Geovane and Vu{\v{c}}ini{\'c}, Mali{\v{s}}a and Watteyne, Thomas},
  booktitle={2025 IEEE Conference on Standards for Communications and Networking (CSCN)},
  pages={1--4},
  year={2025},
  organization={IEEE}
}

@inproceedings{sridhara2024acai,
  title={{ACAI}: Protecting accelerator execution with arm confidential computing architecture},
  author={Sridhara, Supraja and Bertschi, Andrin and Schl{\"u}ter, Benedict and Kuhne, Mark and Aliberti, Fabio and Shinde, Shweta},
  booktitle={33rd USENIX Security Symposium (USENIX Security 24)},
  pages={3423--3440},
  year={2024}
}

@inproceedings{tan2025pipellm,
  title={Pipellm: Fast and confidential large language model services with speculative pipelined encryption},
  author={Tan, Yifan and Tan, Cheng and Mi, Zeyu and Chen, Haibo},
  booktitle={Proceedings of the 30th ACM International Conference on Architectural Support for Programming Languages and Operating Systems, Volume 1},
  pages={843--857},
  year={2025}
}

@inproceedings{wang2024cage,
  title={CAGE: Complementing Arm CCA with GPU Extensions.},
  author={Wang, Chenxu and Zhang, Fengwei and Deng, Yunjie and Leach, Kevin and Cao, Jiannong and Ning, Zhenyu and Yan, Shoumeng and He, Zhengyu},
  booktitle={NDSS},
  year={2024}
}

@article{wang2025building,
  title={Building Confidential Accelerator Computing Environment for Arm CCA},
  author={Wang, Chenxu and Lu, Kun and Zhang, Fengwei and Deng, Yunjie and Leach, Kevin and Cao, Jiannong and Ning, Zhenyu and Yan, Shoumeng and Wei, Tao and He, Zhengyu},
  journal={IEEE Transactions on Dependable and Secure Computing},
  year={2025},
  publisher={IEEE}
}

@inproceedings{walther2022ratls,
  title={RATLS: Integrating transport layer security with remote attestation},
  author={Walther, Robert and Weinhold, Carsten and Roitzsch, Michael},
  booktitle={International Conference on Applied Cryptography and Network Security},
  pages={361--379},
  year={2022},
  organization={Springer}
}

@inproceedings{zhang2023shelter,
  title={{SHELTER}: extending arm {CCA} with isolation in user space},
  author={Zhang, Yiming and Hu, Yuxin and Ning, Zhenyu and Zhang, Fengwei and Luo, Xiapu and Huang, Haoyang and Yan, Shoumeng and He, Zhengyu},
  booktitle={32nd USENIX Security Symposium (USENIX Security 23)},
  pages={6257--6274},
  year={2023}
}

@article{zhang2025complementing,
  title={Complementing Confidential Computing Environment for Applications on Arm CCA},
  author={Zhang, Yiming and Hu, Yuxin and Ning, Zhenyu and Zhang, Fengwei and Luo, Xiapu and Huang, Haoyang and Yan, Shoumeng and He, Zhengyu},
  journal={IEEE Transactions on Dependable and Secure Computing},
  year={2025},
  publisher={IEEE}
}

@inproceedings{zhou2025rcontainer,
  title={RContainer: A Secure Container Architecture through Extending ARM CCA Hardware Primitives.},
  author={Zhou, Qihang and Cao, Wenzhuo and Jia, Xiaoqi and Liu, Peng and Zhang, Shengzhi and Chen, Jiayun and Xu, Shaowen and Song, Zhenyu},
  booktitle={NDSS},
  year={2025}
}

@article{he2025artificial,
  title={Artificial intelligence security and privacy: a survey},
  author={He, Xinlei and Xu, Guowen and Han, Xingshuo and Wang, Qian and Zhao, Lingchen and Shen, Chao and Lin, Chenhao and Zhao, Zhengyu and Li, Qian and Yang, Le and others},
  journal={Science China Information Sciences},
  volume={68},
  number={8},
  pages={181101},
  year={2025},
  publisher={Springer}
}

@manual{arm2024rmm,
  author={{Arm Limited}},
  title={{Realm Management Monitor Specification}},
  organization={Arm Limited},
  year={2024},
  number={DEN0137},
  note={Version 1.0-rel0. Accessed: 2026-05-06},
  url={https://developer.arm.com/documentation/den0137/latest/}
}

@misc{volatility,
  author={{Volatility Foundation}},
  title={The {Volatility} Framework: Volatile Memory Artifact Extraction Utility Framework},
  howpublished={\url{https://www.volatilityfoundation.org/}},
  year={2024},
  note={Accessed: 2026-05-07}
}

@misc{tfrmm,
  author={{Trusted Firmware Project}},
  title={{TF-RMM}: Trusted Firmware Realm Management Monitor},
  howpublished={\url{https://www.trustedfirmware.org/projects/tf-rmm/}},
  year={2024},
  note={Accessed: 2026-05-07}
}

@misc{tfa,
  author={{Trusted Firmware Project}},
  title={{Trusted Firmware-A}},
  howpublished={\url{https://www.trustedfirmware.org/projects/tf-a/}},
  year={2024},
  note={Accessed: 2026-05-07}
}

@misc{linuxcca,
  author={{Arm Limited}},
  title={{linux-cca}: Linux kernel Arm CCA host support branch},
  howpublished={\url{https://gitlab.arm.com/linux-arm/linux-cca}},
  note={Branch: cca-host/v8. Accessed: 2026-05-07}
}

@misc{kvmtool,
  author={{kvmtool contributors}},
  title={{kvmtool}: A lightweight tool for hosting KVM guests},
  howpublished={\url{https://git.kernel.org/pub/scm/linux/kernel/git/will/kvmtool.git}},
  year={2024},
  note={Accessed: 2026-05-07}
}

@misc{cca3world,
  author={{Arm Limited}},
  title={Shrinkwrap cca-3world configuration: Arm {CCA} three-world reference stack},
  howpublished={\url{https://shrinkwrap.docs.arm.com/en/latest/userguide/configstore/cca-3world.html}},
  year={2024},
  note={Accessed: 2026-05-07}
}

@inproceedings{jiang2007stealthy,
  author={Jiang, Xuxian and Wang, Xinyuan and Xu, Dongyan},
  title={Stealthy Malware Detection Through {VMM}-Based ``Out-of-the-Box'' Semantic View Reconstruction},
  booktitle={Proceedings of the 14th ACM Conference on Computer and Communications Security (CCS)},
  pages={128--138},
  year={2007}
}

@inproceedings{payne2008lares,
  author={Payne, Bryan D. and Carbone, Martim and Sharif, Monirul and Lee, Wenke},
  title={Lares: An Architecture for Secure Active Monitoring Using Virtualization},
  booktitle={Proceedings of the 2008 IEEE Symposium on Security and Privacy (S\&P)},
  pages={233--247},
  year={2008}
}

@inproceedings{srinivasan2011process,
  author={Srinivasan, Deepa and Wang, Zhi and Jiang, Xuxian and Xu, Dongyan},
  title={Process Out-Grafting: An Efficient ``Out-of-{VM}'' Approach for Fine-Grained Process Execution Monitoring},
  booktitle={Proceedings of the 18th ACM Conference on Computer and Communications Security (CCS)},
  pages={363--374},
  year={2011}
}
%


\end{document}